# Towards Spatio-Temporal SOLAP

Pablo Bisceglia[1] and Leticia Gómez[2] and Alejandro Vaisman[3]

[1] Universidad de Buenos Aires, Argentina `pbisceglia@gmail.com`
[2] Instituto Tecnológico de Buenos Aires, Argentina `lgomez@itba.edu.ar`
[3] Universidad de la República, Uruguay `avaisman@fing.edu.uy`

**Abstract.** The integration of Geographic Information Systems (GIS) and On-Line Analytical Processing (OLAP), denoted SOLAP, is aimed at exploring and analyzing spatial data. In real-world SOLAP applications, spatial and non-spatial data are subject to changes. In this paper we present a temporal query language for SOLAP, called TPiet-QL, supporting so-called discrete changes (for example, in land use or cadastral applications there are situations where parcels are merged or split). TPiet-QL allows expressing integrated GIS-OLAP queries in an scenario where spatial objects change across time.

## 1 Introduction

In Geographic Information Systems (GIS), spatial data are organized in *thematic layers*, stored in suitable data structures, while associated attributes are usually stored in conventional relational databases. In real-world applications, spatial objects in a layer can be added, removed, split, merged, or their shape may change. Tryfona and Jensen [1] classify spatio-temporal applications according with the kind of support of the changes occurring in the spatial objects. They distinguish between objects with *continuous motion* (e.g., a car moving in a highway), objects with *discrete changes* (e.g, parcels changing boundaries), and objects combining *continuous motion and changing shapes* (e.g., a stain in a river). On the other hand, OLAP (On-Line Analytical Processing) [2] provides a set of tools and algorithms that allow efficiently querying multidimensional repositories called Data Warehouses. OLAP data are organized as a set of *dimension hierarchies* and *fact tables*, and can be perceived as a *data cube*, where each cell contains a measure or set of measures of interest. The problem of integrating OLAP and GIS systems for decision-making analysis has been called SOLAP [3]. One of the models proposed for SOLAP is Piet [4], a framework that integrates spatial, spatio-temporal, and non-spatial multidimensional data. In this paper we add temporal capabilities to SOLAP, extending Piet-QL (the query language associated to the Piet data model) to support *discrete changes.*

*A Motivating Example.* We present a typical scenario about land property information. Figure 1 (left) shows four parcels of land, P1 through P4, probably characterized by attributes like type of soil, owner. We assume that parcels are represented in a GIS layer denoted $L_{land}$. Non-spatial information is stored in a conventional data warehouse. A dimension hierarchy denoted *Land* stores information related to the parcels. The bottom level of this dimension contains the parcel identifiers (p1 through p4). There is a mapping (not shown in the figure) between spatial objects in $L_{land}$ and members of the bottom level (*parcelId*) of the dimension *Land*. At a certain moment, parcels P3 and P4 are merged into a single one $P_{3-4}$. Changes must also be performed at the data warehouse, meaning that elements p3 and p4 are deleted and $p_{3-4}$ is added, along with the corresponding rollups to region r2. A mapping between $p_{3-4}$ and $P_{3-4}$ is also defined. This is depicted on the right hand side of Figure 1. Other changes may also occur. In a *discrete changes* scenario like this, we may want to know the history of $P_{3-4}$, the production of each existing parcel as of the year 2006, or to pose queries like "Production by year per square mile for each parcel of land, for the parcels in Montevideo". Answering these kinds of queries requires extending non-temporal SOLAP data models and query languages (like Piet-QL) with temporal capabilities. This is the problem we address in this paper where, after an overview of related work (Section 2), we define the temporal data model (Section 3). Then (Section 4) we present the syntax and semantics of TPiet-QL, and discuss the expressiveness of the language. We conclude in Section 5.

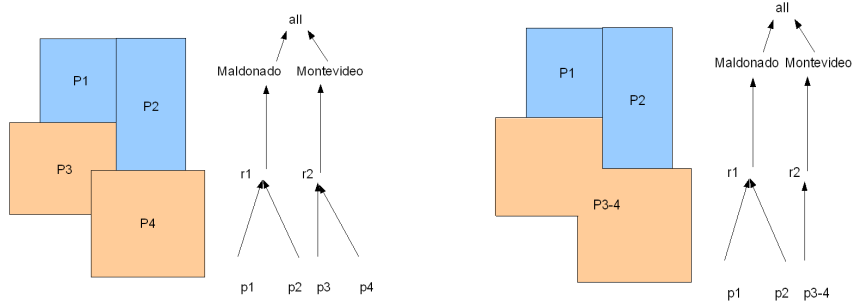

**Fig. 1.** Initial situation (left): land partition and Land dimension hierarchy; after merging P3 and P4 (right): changes in spatial objects and in the dimension hierarchy.

## 2 Related Work

Rivest *et al.* [5] introduced the concept of SOLAP (standing for Spatial OLAP), a paradigm aimed at exploring spatial data by drilling on maps in a way analogous to what is performed in OLAP with tables and charts. Piet [4] is a formal model for SOLAP, where the integration between GIS and OLAP is materialized through a function that maps elements in the data warehouse to elements in the GIS layers. Piet comes equipped with a query language, Piet-QL [6], that supports the operators proposed by the Open Geospatial Consortium[4] for SQL, adding the necessary syntax to integrate OLAP operations through MDX[5]. Piet-QL is designed to support (besides standard GIS and OLAP queries, GIS queries filtered using OLAP conditions, like "Name of the cities with total sales higher that 5000 units"; (d) OLAP queries filtered by spatial conditions, like "Total sales in cities within 100Km from Montevideo". Filtering is implemented through a predicate denoted IN. The Piet-QL query "Parcels crossed by the 'Uruguay' river, with sales greater than 5000 units" reads in Piet-QL.

```
SELECT GIS l.id
FROM land l, rivers lr
WHERE intersects(l,lr) AND lr.name = "Uruguay" AND l IN(
    SELECT CUBE filter([Land].[Land parcelId].Members,
    [Measures].[Parcel Sales] > 5000)
    FROM [Sales]);
```

Here, 'land' and 'rivers' represent two thematic layers containing spatial objects (the parcel subdivision of a given region, and the rivers, respectively). The OLAP subquery (identified with the keyword CUBE) is linked to the outer query by the predicate IN, and returns a collection of identifiers of spatial objects.

The Spatio-Temporal Relational data Model (STRM), introduced by Tryfona and Hadzilacos [7], provides a set of constructs consisting in relations, layers, virtual layers, object classes, and constraints, all with spatial and temporal extent, on top of well-known models. In this model, a *layer* is a set of geometric figures like points, lines, regions or combinations of them, with associated values. The authors also define a layer algebra, which, based on four operations over layers, provides a semantics to SOLAP.

Other proposals such as SECONDO [8] and Hermes [9] support moving object databases but, like other spatio-temporal models (except Piet), they are not oriented towards addressing the problem of integrating GIS, OLAP and Moving Object data.

## 3  Spatio-Temporal Piet

In the temporal extension to Piet (TPiet), each tuple in a relation is timestamped with its validity interval. Time is introduced as a new sort (domain). For clarity of presentation, in the sequel we work with point-based temporal domains, although we use interval-based domains to implement our ideas [10]. In temporal databases, the concepts of *valid* and *transaction* times refer to the instants when data are valid in the real world, and when data are recorded in the database, respectively [11]. We assume *valid* time support. Also, a distinguished variable *Now* represents the (moving) current time instant. The *lifespan* of a GIS layer $L$, $lifespan(L)$, is the collection of all the time instants where the layer is valid. The *lifespan* of a set of layers $\mathcal{L}$, $lifespan(\mathcal{L})$, is the union of the lifespans of all the layers in $\mathcal{L}$. Finally, we assume that no structural changes occur at the GIS or at the data warehouses, meaning that a layer containing polygons at its creation instant will contain polygons throughout its lifespan.

---

[4] http://www.opengeospatial.org
[5] http://msdn2.microsoft.com/en-us/library/ms145506.aspx.

Given the above, a *Temporal GIS-OLAP Dimension Schema* $TG_{sch}$ is a tuple $\langle H, \mathcal{A}, \mathcal{D}, \mu \rangle$, where $H$ is a mapping from layers to geometries, $\mathcal{A}$ is a set of *partial* functions $Att$ that map attributes in OLAP dimensions to GIS layers, $\mathcal{D}$ is a set of dimension hierarchies [12], and $\mu$ a dimension level in a standard OLAP Time dimension. Elements in $\mu$ are in the temporal domain. Further, $H$, $\mathcal{A}$, and $\mathcal{D}$ satisfy the following conditions: (a) A layer is created when the first object is added to it; (b) $H$ is constant throughout the lifespan of the GIS; (c) For each layer $L$, the function $Att$ is defined only in $lifespan(L)$; (d) The functions $Att \in \mathcal{A}$ do not change with time, i.e., $Att_1(parcelId, Land)$ will always return $L_{land}$. (e) The schema of the dimensions in $\mathcal{D}$ is constant during the lifespan of the GIS. Associated with a dimension schema, we have a dimension instance, which consists in: A set of relations $r_{L_i}^t$ such that each tuple $\langle g_i, ext(g_i), t \rangle$ in $r_{L_i}^t$, represents the existence of an object $g_i$ (and its extension) in $L_i$ at the instant $t$; A collection of functions $\alpha$ that map elements in OLAP dimension levels to geometric elements in a GIS layer, at a given time; A collection of dimension instances, one for each dimension schema $D \in \mathcal{D}$ in $TG_{sch}$. We assume that spatial objects have the same attributes throughout their lifespan.

*Temporal Piet Data Structure* The data structure of TPiet-QL is organized in: (a) Application information. This is the data warehouse structure. Contains dimension and fact tables. (b) GIS information. The data structure for the map layers (one table per layer). Temporal attributes FROM and TO indicate the interval of validity of each object in a layer. (c) GIS-OLAP mapping information. Stores the relationship between geometric and application information (i.e., the $\alpha$ functions). Temporal attributes are also included here to indicate the temporal validity of a mapping. (d) There are also data structures to store precomputed information containing the overlay of different layers (see [4]).

We briefly explain the update semantics. When a new object is *created* at instant $t_1$, say, in the layer *Land*, a tuple is inserted in the *Land* table with the corresponding parcel information. Attributes FROM and TO are set to $t_1$ and the distinguished value *Now*, respectively. If this parcel, call it $p_1$, is *split* into $p_2$ and $p_3$ at instant $t_2$, the tuple for $p_1$ is timestamped with TO=$t_2 - 1$ (i.e., an instant immediately before $t_2$ in the object's granularity); in addition, two tuples are *created* for $p_2$ and $p_3$, with FROM=$t_2$, and TO=*Now*. Later, at $t_4$, two parcels, $p_5$ and $p_6$ are *merged* into a single one, call it $p_{56}$. The former two tuples are *deleted* as before (i.e., timestamped with TO=$t_4 - 1$), and $p_{56}$ is *created* with FROM=$t_4$ and TO=*Now*. The *update* operation at instant $t$ is equivalent to the *deletion* of a tuple (i.e., a timestamping with $t - 1$), and the *insertion*, at instant $t$, of a new one (keeping the same identifier). The *reincarnation* operator is analogous to an update, except for the fact that the instants of deletion and insertion are not consecutive.

We now discuss the *data warehouse* side. When operations on the GIS side require creating new spatial objects, the corresponding objects must be inserted in the warehouse dimensions, also defining new mappings. However, when an *update* occurs (like a change in an object's shape) the object identifier does not change and no action needs to be taken on the warehouse side. Also note that insertions can be performed without impacting the warehouse or the mapping function, although this could produce incomplete answers to some queries (the ones that involve accessing the warehouse), due to the incomplete mapping (i.e., the object would only be in one of the parts of the system). One of the premises of the Piet data model is to allow autonomous maintenance of warehouse and GIS information. There are at least two possible situations: (a) The data warehouse and associated data cubes are non-temporal, in the sense that only fact tables are updated, and the dimensions are *static*, i.e., only the current state of the dimension data is available; (b) The data warehouse has *temporal* capabilities, i.e., dimensions are updated and their history is preserved. For example, the notion of *slowly changing dimensions* can be used [2], where a new dimension tuple is added when an update occurs (dimension tables are extended with FROM/TO attributes). Other solutions can be found in the literature [13,14].

## 4 Query language

**Definition 1 (Spatio-temporal object).** *We denote by spatio-temporal object a tuple of the form $\langle objectId, geometry, attribute_1, ..., attribute_n, interval \rangle$, where geometry is the geometric extension of the object, $attribute_i$ are alphanumeric attributes, and 'interval' is the interval of validity of the object, of the form $[FROM, TO]$.* □

In Definition 1, *interval* is a single interval. In temporal databases it is usual to talk about temporal elements, i.e., sets of intervals. For simplicity of presentation, in this paper we work with single intervals instead of temporal elements. This makes the paper easier to read, without reducing its substance. In what follows we refer to spatio-temporal objects as 'objects', and denote $\mathcal{G}$ a collection of spatio-temporal objects. Based on Allen's interval set of predicates [15], in Figure 2 we specify the syntax and semantics of a collection of predicates over spatio-temporal objects, intervals, and time instants.

Note that `DURING` and `COVERS` represent the predicate `X DURING Y` in Allen's algebra. `OVERLAPS` represents `X OVERLAPS Y` and `Y OVERLAPS X`. The same for `MEETS`, `STARTS,` and `FINISHES`. `BEFORE` and `AFTER` represent X < Y and Y < X, respectively. We also need some functions, namely:

`IIntersection`$(I_1, I_2)$: $T \times T \times T \times T \to T \times T$; returns the interval when $I_1$ and $I_2$ intersect.

`Coalesce`$(\mathcal{G})$: Analogously to the 'Coalesce' operator used in temporal databases, it produces groups of objects whose temporal intervals are consecutive and that coincide in all other attributes, returning a collection of spatio-temporal objects.

| | |
|---|---|
| `StartsBefore(g,t)`: $\mathcal{G} \times T \to boolean$; Given a spatio-temporal object and an instant, returns *True* if $t > $ `g.FROM`. | `FinishesAfter(g,t)`: $\mathcal{G} \times T \to boolean$; Given a spatio-temporal object and an instant, returns *True* if $t < $ `g.TO`. |
| `BeginsAfter (g,t)`: $\mathcal{G} \times T \to boolean$; Given a spatio-temporal object and an instant, returns *True* if $t < $ `g.FROM`. | `AT(g,t)`: $\mathcal{G} \times T \to boolean$; Given a spatio-temporal object and an instant, returns *True* if $t_1 \leq $ `g.FROM` AND $t_1 >= $ `g.TO`. |
| `BEFORE` $(g, \langle t_1, t_2 \rangle)$: $\mathcal{G} \times T \times T \to boolean$; Given a spatio-temporal object and an interval, returns *True* if `g.TO` $< t_1$. | `AFTER`$(g, \langle t_1, t_2 \rangle)$: $\mathcal{G} \times T \times T \to boolean$; Given a spatio-temporal object and an interval, returns *True* if $t_2 <$ `g.FROM`. |
| `DURING`$(g, \langle t_1, t_2 \rangle)$: $\mathcal{G} \times T \times T \to boolean$; Given a spatio-temporal object and an interval, returns *True* if $t_1 \leq $ `g.FROM` AND $t_2 \geq $ `g.TO`. | `OVERLAPS`$(g, \langle t_1, t_2 \rangle)$: $\mathcal{G} \times T \times T \to boolean$; Given a spatio-temporal object and an interval, returns *True* if $(t_1 <$ `g.FROM` AND $t_2 >$ `g.FROM` AND $t_2 <$ `g.TO`) OR $(t_1 >$ `g.FROM` AND $t_2 >$ `g.TO` AND $t_1 <$ `g.TO`). |
| `COVERS`$(g, \langle t_1, t_2 \rangle)$: $\mathcal{G} \times T \times T \to boolean$; Given a spatio-temporal object and an interval, returns *True* if $t_1 \geq $ `g.FROM` AND $t_2 \leq $ `g.TO`. | `MEETS`$(g, \langle t_1, t_2 \rangle)$: $\mathcal{G} \times T \times T \to boolean$; Given a spatio-temporal object and an interval, returns *True* if $t_1 = $ `g.TO` OR $t_2 = $ `g.FROM`. |

**Fig. 2.** Predicates over spatio-temporal objects, intervals, and instants.

*Spatio-temporal Joins* A key operation in any spatio-temporal query language is the *join*. Different kinds of temporal joins have been proposed in the literature [11], and two main classes can be identified: (a) Disjoint join; and (b) Overlap join. In the former, given $n$ (timestamped) tuples, it is not required that their time intervals overlap. In the latter, the time intervals must overlap and there are two possibilities: all the time intervals have at least one common time instant, or they are joined in a 'chained' fashion, e.g., $t_1.TO \geq t_2.FROM \wedge t_2.TO \geq t_1.TO$. Disjoint joins provide more expressiveness to a query language than overlap joins, allowing to query for asynchronous events (e.g., parcels owned by X before a region changed name). Examples (following Allen [15]) are `before-join(X,Y)`, and `meet-join(X,Y)`, with conditions $X.TO \leq Y.FROM$ and $X.TO = Y.FROM$, respectively. The joins above are denoted T-joins. When a T-join requires the equality of a collection of non-temporal attributes specified as a predicate $P_a$, we say that we are in presence of a GT-join (standing for generic temporal). That is, a GT-join corresponds to the expression $\sigma_{P_a \wedge overlap-join(X,Y)}(X, Y)$. That means, given two tuples, the tuples in the result of a GT-join will be the ones that have overlapping time intervals and verify the non-temporal predicate $P_a$. In a spatio-temporal setting we can implement the temporal joins using the operators defined above.

In the presence of spatio-temporal objects, the GT-join can be defined using the standard topological relationships [16], like `Touches`$(g_1, g_2)$, or `Contains`$(g_1, g_2)$. Consider two layers storing the histories of airports and cities. Figure 3 (left) shows two stages of city $c_1$: one in the interval [0,50], and the other in the interval [51,Now]. Airport $a_1$ was first relocated at instant 100, and then, due to the city expansion, it was located well outside the new city limits. Figure 3 (right) shows how the two objects $a_1$ and $c_1$ interact along their timelines: the airport is within the city limits only in the intervals [51,100] and [101,200]. The relational representations are given below.

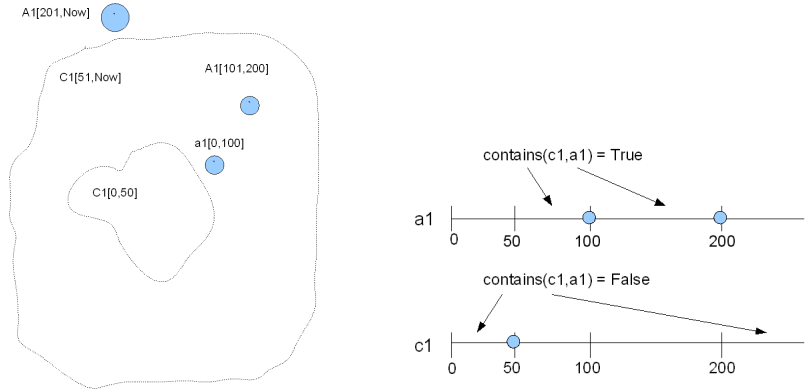

**Fig. 3.** A city an its airport (left); Interaction of $a_1$ and $c_1$ along their timelines (right)

| cityId | the_geom | ... | FROM | TO  |
|--------|----------|-----|------|-----|
| c1     | g1       | ... | 0    | 50  |
| c1     | g2       | ... | 51   | Now |
| c2     | g3       | ... | 0    | 30  |

| airportId | the_geom | ... | FROM | TO  |
|-----------|----------|-----|------|-----|
| a1        | g1       | ... | 0    | 100 |
| a1        | g2       | ... | 101  | 200 |
| ...       | ...      | ... | ...  | ... |

We can list the pairs city-airport such that an airport was within the city limits as a GT-join, where the non-temporal predicate *Contains* is spatial:

$\sigma_\phi(Airports \times Cities)$
$\phi = contains(Airports.geom, Cities.geom) \wedge overlap-join(Airports, Cities)$

The result would contain the tuples $\langle a1, c1, 51, 100 \rangle$ and $\langle a1, c1, 101, 200 \rangle$, representing (see Figure 3), that between instants 51 and 100, a1 remained within the city limits of c1.

*The TPiet-QL Query Language* The discussion above set the basis for defining a temporal extension to the GIS part of Piet-QL, yelding the TPiet-QL language.

SELECT GIS [SNAPSHOT][CURRENT] list_of_attributes
FROM [OVERLAP] T1 t1,...,Tn tn
WHERE $\Phi$

T1 through Tn represent thematic layers, t1 through tn range over the spatial or *spatiotemporal objects* in these layers, and the $a_i$'s represent attributes of these objects. The OVERLAP keyword in the FROM clause states that the overlap join semantics must be applied (see below). The list of attributes in the SELECT clause defines the schema of the result: a subset of the union of the attributes of the spatiotemporal objects mentioned in the FROM clause. The SNAPSHOT keyword (analogous to the one in TSQL2 [17]) is used to return a non-temporal relation, eliminating the interval/s associated with each tuple in the query result. CURRENT is the same as SNAPSHOT but selecting the *current* state of the relation before the projection is performed. That means, the query will return a collection of spatial objects corresponding to the spatiotemporal ones which contain the keyword 'Now' in the attribute TO.

The condition $\Phi$ is composed of conjunctions and disjunctions of the function and predicates mentioned above, and can also include the Piet-QL predicate IN (and the corresponding OLAP sub-query), to provide compatibility with Piet-QL, and to support OLAP in a spatio-temporal SOLAP scenario. This is why we keep the Piet-QL keyword GIS in the SELECT clause. We show this below by means of some examples.

The semantics of the query is defined by the cartesian product of the geometric objects in all the thematic layers in the FROM clause. If the OVERLAP keyword is specified, only the tuples whose intervals overlap are considered, (ie., the tuples such that $\cap_{ti.interval, i=1,n} \neq \emptyset$), and the overlapping interval is included in the result, which is coalesced by default using all the non-temporal attributes in the SELECT clause. We illustrate this semantics extending the city-airport example with a layer containing parcels, described in the table below (on the right we show the distances between cities and parcels, during different time intervals):

| parcelId | the_geom | ... | FROM | TO  |
|----------|----------|-----|------|-----|
| p1       | g1       | ... | 10   | 20  |
| p1       | g2       | ... | 21   | 40  |
| p2       | g3       | ... | 30   | 50  |
| p3       | g4       | ... | 40   | 100 |
| ...      | ...      | ... | ...  | ... |

| cityId | parcelId | FROM | TO  | distance |
|--------|----------|------|-----|----------|
| c1     | p1       | 10   | 20  | 80       |
| c1     | p1       | 21   | 40  | 120      |
| c1     | p2       | 30   | 50  | 70       |
| c1     | p3       | 40   | 50  | 80       |
| c1     | p3       | 51   | 100 | 90       |

Consider a query asking for pairs city-parcel such that the distance between them is/was less than 100Km. According to the usual semantics of a temporal join, the query returns tuples of the form $\langle p_i, c_j, Interval \rangle$, where $Interval$ is the interval when they where closer than 100Km from each other. The query reads in TPiet-QL:

SELECT GIS c,p
FROM OVERLAP Parcels p, Cities c
WHERE Distance(c.the_geom,p.the_geom) < 100

The result will be (note that this result is coalesced):

| cityId | parcelId | FROM | TO  |
|--------|----------|------|-----|
| c1     | p1       | 10   | 20  |
| c1     | p2       | 30   | 50  |
| c1     | p3       | 40   | 100 |

Let us give now an example of a TPiet-QL query returning an OLAP cube filtered with a spatio-temporal sub-query containing with SNAPSHOT clause: "Production cost and parcel sales in 2009, for the parcels crossed by rivers at that time". This query reads:

    SELECT CUBE [Measures].[Production Cost], [Measures].[Parcel Sales],
    Product.[All_Products] ON ROWS
    FROM [Sales]
    WHERE AND [Time].[2009] AND
    [Land].[All Land] IN (
      SELECT GIS SNAPSHOT l.id
      FROM OVERLAP Land l, Rivers r
      WHERE Crosses(r,l) AND
      COVERS(r,[1/1/2009,12/31/2009]) AND
      COVERS(l,[1/1/2009,12/31/2009]) ) ;

We conclude with the query: "Parcels crossed by the Uruguay river, with production sales greater than 5000 units". (Technically, in TPiet-QL, a GIS-OLAP query).

    SELECT GIS l
    FROM OVERLAP land l, rivers r
    WHERE Crosses(l,r) AND r.name = "Uruguay" AND l.id IN(
      SELECT CUBE
      filter([Land].[Land parcelId].Members,
      [Measures].[Parcel Sales] > 5000)
      FROM [Sales]);

The query returns the spatiotemporal objects containing the parcels with the requested production, their information, and the intervals when each parcel in the result crossed the Uruguay river.

*Expressive Power* Over the data model described in Section 3, a formal spatio-temporal query language, denoted $\mathcal{L}_t$, has been defined. This query language is studied in detail in [18]. We show now that TPiet-QL is based on this formal query language, and that most queries expressible in $\mathcal{L}_t$ are captured by this temporal extension to Piet-QL. We illustrate these ideas using a very simple GIS-OLAP query, which includes a reference to an external data cube called 'Production', with dimensions Land and Time, and measure 'quantity', representing the production of wheat per year. The query asks for the parcels having an area larger than 100 Ha in 1996, currently larger than they were at that time, and with a production of wheat larger than 1000 Tons in 2009. The formal query in $\mathcal{L}_t$ reads:

$$Q = \{p \mid (\exists e_p)(\ \exists e_{p_1})(\exists a)(\exists qty)$$
$$(r^t_{L_{land}}(p, e_{p_1}, 1996) \ \wedge \ r^t_{L_{land}}(p, e_p, Now) \ \wedge$$

$$area(e_{p_1}) = a \ \wedge \ a > 100 \ \wedge \ area(e_p) > a) \ \wedge$$
$$Production(p, 2009, qty) \ \wedge \ qty > 1000\}.$$

Here, $Production(p, 2009, qty)$ is a term representing a fact table, $area$ is a function computing the area of a spatial object, and $r^t_{L_{land}}(p, e_p, t)$ are terms representing the parcels and their geometric extensions across time (in a point-based fashion), corresponding to the elements in the model of Section 3. This query can be expressed in TPiet-QL as follows:

```
SELECT GIS p1.id
FROM land p, land p1
WHERE area(p) > area(p1) AND
COVERS(p,[1996,1996]) AND COVERS(p1,[2010,2010]) AND
p1.id=p.id AND p1.id IN(
    SELECT CUBE
    filter([Land].[Land parcelId].Members,
    [Measures].[qty] > 1000)
    FROM [Production])    SLICE [Time].[2009];
```

The constructs $\mathcal{L}_t$ are present in the TPiet-QL expression above. The main difference is that instead of using non-temporal functions over the extensions of spatial objects like in $\mathcal{L}_t$, i.e., while $area$ is applied over a geometry (e.g., $e_p$), TPiet-QL uses temporal functions over spatio-temporal objects (e.g., $p$). It can be seen that queries expressible in $\mathcal{L}_t$ can be expressed in TPiet-QL since there is a translation for each of the terms in one language to the other. We omit a term-by term proof for the sake of space.

Vaisman and Zimányi [19] recently proposed a comprehensive and formal classification for spatio-temporal data warehousing, defining a taxonomy of queries. For example, the SOLAP class of queries is defined as the class containing the queries that can be expressed in relational calculus with aggregate functions, extended with *spatial* data types. Analogously, the ST-OLAP class of queries is the class containing the queries that can be expressed in the calculus extended with spatial and the *moving* types defined in [20]. We can say that our proposal falls somewhere in between the ST-OLAP and ST-TOLAP classes (the latter includes temporal OLAP support).

## 5 Conclusion and Future Work

We have presented a spatio-temporal query language for temporal SOLAP, denoted TPiet-QL, that supports discrete changes in the spatial objects in the thematic layers of a GIS. TPiet-QLextends Piet-QL, a query language for SOLAP. We introduced the syntax an semantics of the language, and discussed its expressive power. Our next step is to produce an implementation, which includes a visualization tool for spatio-temporal data, and the development of efficient methods for query processing.

**Acknowledgements:** The authors of this paper were partially funded by the LACCIR project "Monitoring Protected Areas using an OLAP-enabled Spatio-temporal GIS".